\documentclass[nofootinbib,10pt, twocolumn,prd,preprintnumbers,superscriptaddress,aps]{revtex4-1}

\usepackage{tabularray}
\usepackage{amsfonts}
\usepackage{graphicx}
\usepackage[colorlinks=true,linkcolor=blue,citecolor=teal,urlcolor=blue]{hyperref}
\usepackage[dvipsnames]{xcolor}
\usepackage{amsmath}
\usepackage{cases}
\usepackage{braket}
\usepackage[T1]{fontenc}
\usepackage{mathrsfs}
\usepackage{enumerate}
\usepackage{bm}
\usepackage{multirow}
\usepackage{comment}
\usepackage{makecell}
\usepackage{subfigure}
\usepackage{aas_macros}
\usepackage{soul}

\begin{document}

\title{Measuring the boson mass of fuzzy dark matter with stellar proper motions}

\author{Gabriel d'Andrade Furlanetto}
\email{gabdandrade@usal.es}
\affiliation{{Departamento de F\'isica Fundamental, Universidad de Salamanca, Plaza de la Merced, s/n, E-37008 Salamanca, Spain}}

\author{Riccardo Della Monica}
\email{rdellamonica@usal.es}
\affiliation{{Departamento de F\'isica Fundamental, Universidad de Salamanca, Plaza de la Merced, s/n, E-37008 Salamanca, Spain}}

\author{Ivan De Martino}
\thanks{Corresponding author}
\email{ivan.demartino@usal.es}
\affiliation{{Departamento de F\'isica Fundamental, Universidad de Salamanca, Plaza de la Merced, s/n, E-37008 Salamanca, Spain}}
\affiliation{{Instituto Universitario de Física Fundamental y Matemáticas, Universidad de Salamanca, Plaza de la Merced, s/n, E-37008 Salamanca, Spain}}

\begin{abstract}
Fuzzy Dark Matter (FDM)  is among the most suitable candidates to replace WIMPs and to resolve the puzzling mystery of dark matter. A galactic dark matter halo made of these ultralight bosonic particles leads to the formation of a solitonic core surrounded by quantum interference patterns that, on average, reproduce a Navarro-Frenk-White-like mass density profile in the outskirts of the halo. The structure of such a core is determined once the boson mass and the total mass of the halo are set. We investigated the capability of future astrometric Theia-like missions to detect the properties of such a halo within the FDM model, namely the boson mass and the core radius. To this aim, we built mock catalogs containing three-dimensional positions and velocities of stars within a target dwarf galaxy. We exploited these catalogs using a Markov Chain Monte Carlo algorithm and found that measuring the proper motion of at least 2000 stars within the target galaxy, with uncertainty $\sigma_v \leq 3$ km/s on the velocity components, will constrain the boson mass and the core radius with 3\% accuracy. Furthermore, the transition between the solitonic core and the outermost NFW-like density profile could be detected with an uncertainty of 7\%. Such results would not only help to confirm the existence of FDM, but they would also be useful for alleviating the current tension between galactic and cosmological estimations of the boson mass, or demonstrating the need for multiple particles with a broad mass spectrum as naturally arise String Axiverse.

\end{abstract}

\maketitle

\section{Introduction}\label{sec:intro} 

A substantial part of the total amount of matter in the Universe cannot be observed in the electromagnetic spectrum, but its existence manifests itself through gravitational effects. Indeed, an overwhelming amount of evidence, initially from the rotation curves of spiral galaxies \cite{Rubin1970,Rubin1982}, suggested the existence of a non-baryonic and non-luminous constituent whose interaction with the standard model's particles must be extremely weak \cite{bertone_history_2018}. This component is usually called dark matter (DM) and constitutes approximately 84\% of the total matter in the Universe \cite{Planck2020}. {Cosmological observations suggest that the constituent particles of DM must be} massive, non-relativistic, and collisionless. Such a candidate is usually referred to as Cold Dark Matter (CDM). However, observations on galactic scales seem to indicate that CDM faces some issues that are hard to overcome without tuning baryonic feedback \cite{deBlok2010,Boylan2011,Boylan2012,Bullock2017,DeMartino2020,Salucci2021}. In broad strokes, cosmological simulations of CDM show some inconsistencies with observations on small scales. {A well-known example, usually referred to as the `core-cusp problem', is the incompatibility between} numerical N-body simulations of CDM {that} predict a cuspy density profile for halos \cite{Navarro_NFW,de_Blok_CoreCusp}, \emph{i.e.} at small radii $\rho \sim r^\alpha$ with $\alpha=-1$, {and} observations of low surface brightness (LSB) or dwarf spheroidal (dSph) galaxies, in which DM dominates the dynamics of the entire system, {that} show cored mass density profiles \cite{deBlok2001,de_Blok_CoreCusp}. Additionally, simulations predict that the most massive DM halos, such as the one of the Milky Way, should be orbited by small subhalos. This is indeed the case, but the problem appears in the number of subhalos on the lower end of the mass spectrum. The density of these subhalos diverges for small masses, and simulations predict that we should observe several hundred of these small mass populations in the Milky Way. Only a few tens have been observed \cite{Read2019} originating the `missing satellite problem' (for a comprehensive review on the small-scale problems of CDM, we refer to \cite{DeMartino2020} and the references therein).

With this in mind, a question arises: which (possibly fundamental) particle at large scales makes up a fluid that behaves like CDM, is sufficiently abundant to explain the current DM density, and at the same time is capable of solving the issues faced by CDM on galactic scales? Although no particle in the Standard Model (SM) of particle physics fits that description, there is a wide range of promising candidates beyond the SM spanning a wide range of masses such as warm dark matter (WDM) with a mass ranging from 2.5 to 30 keV, QCD axions with a mass of the order of few eV, and ultralight bosons with mass spanning the range $\sim 10^{-24}- 10^{-17}$~eV. Although all of these particles are interesting and have non-trivial and distinct small-scale phenomenology \cite{Khlopov1985,Berezhiani1992,Khlopov1999}, one of the most interesting and intriguing candidates is represented by ultralight bosonic particles with masses $\sim 10^{-22}$ eV, which are usually referred to as Fuzzy Dark Matter (FDM) or Wave Dark Matter ($\psi$DM), and whose de Broglie wavelength is roughly $\sim 1$~kpc (assuming virial velocities for the FDM particles of roughly a few tens km/s) pointing out the emergence of quantum behaviors on astrophysical scales (for a comprehensive review, we refer to \cite{hui_ultralight_2017,ferreira_ultra-light_2021}). 

FDM particles are generated through symmetry breaking via the misalignment mechanism, a phenomenon commonly found in the string landscape \cite{Preskill1983,Abbott1983}. Since FDM particles are spinless, they follow a Bose-Einstein statistic \cite{Sikivie2009} and therefore give rise to a self-gravitating halo sustained by internal quantum pressure due to the uncertainty principle. Indeed, N-body simulations have shown that a FDM halo consists of a solitonic core surrounded by quantum interference patterns on the de Broglie scale whose azimuthal average follows an Navarro-Frenk-White (NFW)-like mass density profile. This implies that the halo mass density distributions is cored rather than cusped in the center. Moreover, still due to the uncertainty principle, FDM cannot form self-gravitating structures on scales smaller than its de Broglie wavelength, thus providing a natural lower Jeans scale for structure formation \cite{schive_cosmic_2014, Davies2020} which would serve to resolve the missing satellite problem. {Among the different possible realizations of FDM \cite{ferreira_ultra-light_2021}, in this study we focus on the simplest possible description, \emph{i.e.} non-interacting FDM \cite{hui_ultralight_2017} falling in the non-relativistic regime. The dynamics of this type of FDM is described by non-linear Schrödinger equation subjected to a gravitational potential, coupled to the Poisson equation.}

{Despite this simple description, non-interacting FDM (hereafter simply FDM)} offers solutions to several small-scale challenges faced by CDM, it has been the subject of several observational tests devoted to constrain the mass of the boson and find its signatures in the data. Starting from small astrophysical scales, an upper limit on the boson mass $m_\varphi \lesssim 3.2\times 10^{-19}$ eV, at the 95 \% confidence level, has been set using observations of the orbital motion of the S2 star around the supermassive black hole in the center of the Milky Way \cite{DellaMonica2023}. Additionally, the gravitational waves emitted from spinning neutron stars at the Galactic Centre can carry out a signature of the existence of those ultralight bosons that third generation detectors such as the Einstein Telescope and the Cosmic Explorer may detect \cite{Blas2024}. On galactic scales, observations of the velocity dispersion profile in dSph galaxies are reproduced with a boson mass of $\sim 10^{-22}$ eV \cite{Chen2017}. Moreover, kinematics of the star in low-density self-gravitating systems such as ultra-diffuse galaxies is also reproduced with a similar boson mass \cite{Broadhurst2020, Pozo2021}. Additionally, a Jeans analysis of local isolated dwarf galaxies also favors a boson mass of the order of $\sim 10^{-22}$ eV to explain the cored DM mass density profile and the transition scale observed in the stellar number density profiles \cite{Pozo2020, Pozo2023}. Finally, 
the same boson mass may serve to explain the velocity dispersion of the stars within the bulge of the Milky Way \cite{DeMartino2020PDU}.  On larger scales, a recent result claims that anomalies and corrugations in gravitational lensing images may be an observational consequence of the quantum interference pattern of the FDM model \cite{Amruth2023}.  

Conversely, cosmological analyses constrain the value of of the mass of the FDM particle in a quite different range. Analyses of the cosmic microwave background power spectrum provide very weak constraints $10^{-24} \lesssim m_\varphi \lesssim 10^{-17}$ \cite{Hlozek2015, Hlozek2018}, while analyses of the Lyman-$\alpha$ forest set lower bounds on the boson mass $\gtrsim 2\times10^{-20}$ eV \cite{Armengaud2017,Rogers2021}.  Moreover, it has been shown that the mass density profile of ultra-faint galaxies requires a boson mass $\gtrsim 10^{-21}$ eV to explain their total mass. Otherwise, using a boson with mass of the order of $\sim 10^{-22}$~eV, as required in dSph galaxies, the mass of ultra-faint galaxies would be overestimated by 5-6 orders of magnitude \cite{Safarzadeh2020}. These challenges, faced by the FDM model, are still open and under debate and cannot even be solved with more complex kinematic models in dSph galaxies, that assume a radially dependent velocity anisotropy parameter \cite{deMartino2023}. Although these problems have not yet been resolved, FDM remains an extremely interesting candidate, whose observational signatures could be searched, for example, in the timing of pulsars near the Galactic center \cite{DeMartino2017} using forthcoming infrastructures such as the Square Kilometer Array (SKA).

In any case, the possibility of distinguishing at the galactic scale between having a boson with mass $m_\varphi\sim 10^{-22}$~eV or $m_\varphi\sim 10^{-20}$~eV, could on the one hand help to resolve the above mentioned tensions and on the other help to confirm or rule out the scenario of the String Axiverse in which many bosons with different masses are produced and, hence, one uses heavier bosons to account for the kinematic of stars in dwarf galaxies and lighter bosons for explaining other observations \cite{Pozo2023}.

For such a reason, we will study the possibility of measuring the boson mass with future astrometric measurements of the velocity field in dSph galaxies and will estimate the precision down to which the boson mass might be constrained. We aim to demonstrate that  using proper motion measurements, instead of the line-of-sight velocity dispersion, we might constrain the boson mass with an accuracy relatively small to undoubtedly select its value on galactic scales. In Section \ref{sect:fdm}, we summarize the main features of FDM and introduce the main equations to describe their cosmological evolution and the small-scale distribution. In Section \ref{sect:methodology}, we describe the methodology that we adopt to forecast the accuracy achievable with next-generation astrometric missions on the boson mass and core radius parameters, and we explain how the DM halo is built, how stars are distributed,  and how we build the mock catalogs containing the position and velocity components of each star. We also explain the statistical approach used in our forecast analysis, based in a Markov Chain Monte Carlo (MCMC) algorithm. Finally, in Sections \ref{sect:results} and \ref{sect:conclusions} we show our results and give the main conclusions, respectively.

\section{A summary of Fuzzy Dark Matter}\label{sect:fdm}

The dynamics of ultralight bosons is described in terms of a scalar field, $\varphi,$ minimally coupled to the space-time metric\footnote{Here, we are neglecting self-interaction and coupling to ordinary matter.} ($g_{\mu\nu}$) \cite{Hui2017}. In the most general case, one considers a complex scalar field whose dynamics is regulated by the following action (using natural units $G=c=\hbar=1$)
\begin{equation}
    \mathcal{S}_\varphi = \int d^4x\sqrt{-g}\left(-g^{\mu\nu}\nabla_\mu\varphi^*\nabla_\nu\varphi-m_\varphi^2\varphi^*\varphi\right).
    \label{eq:scalar_field_action}
\end{equation}
with $\nabla_\mu$ being the covariant derivative related to the metric $g_{\mu\nu}$ and adopting the spacetime signature ($-$, $+$, $+$, $+$). Since the dynamics of the gravitational field is regulated by the Hilbert-Einstein action, $\mathcal{S}_\textrm{GR}$, the dynamics of a self-gravitating scalar field will be governed by the action $\mathcal{S}=\mathcal{S}_\textrm{GR} + \mathcal{S}_\varphi$ whose variation with respect to $g_{\mu\nu}$ and $\varphi$ returns the following set of fully relativistic non-linear equations,
\begin{align}
    &\frac{G_{\mu\nu}}{8\pi} = \left(\partial_{(\mu}\varphi^*\partial_{\nu)}\varphi - g_{\mu\nu}\left(\frac{1}{2}g^{\alpha\beta}\partial_\alpha\varphi^*\partial_\beta\varphi + m_\varphi^2\varphi^*\varphi\right)\right),\label{eq:field_equation_scalar}\\
    &(\Box - m_\varphi^2)\varphi = 0,
    \label{eq:klein_gordon}
\end{align}
where the D'Alambertian is defined as
\begin{equation}
    \Box\varphi = g^{\mu\nu}\nabla_\mu\nabla_\nu\varphi = \frac{1}{\sqrt{-g}}\partial_\mu(g^{\mu\nu}\sqrt{-g}\partial_\nu\varphi).
\end{equation}

For our purposes, we will restrict to the case of a real scalar field $\varphi$ of mass $m_\varphi$, and focus on its dynamics in the non-relativistic regime which is relevant for structure formation. Therefore, a self-gravitating system composed of a large collection of bosonic particles, sharing the same quantum state, will form a Bose-Einstein condensate on galactic scales that compose the FDM halo. Under these assumptions, the scalar field $\varphi$ can be written as
\begin{equation}
    \varphi(t,\vec{r}) = \frac{1}{\sqrt{2m_\varphi}}[\psi(t,\vec{r})e^{-im_\varphi t}+\psi^*(t,\vec{r})e^{im_\varphi t}].
\end{equation}
Here the time dependence of $\varphi$ over the scalar field Compton frequency has been factored out, and the wavefunction  $\psi$ has been introduced. In the non-relativistic regime, the field equations in equation \eqref{eq:field_equation_scalar} reduce to those of  Newtonian gravity, while the Klein-Gordon equation in equation \eqref{eq:klein_gordon} reduces to the Schrödinger-Poisson (SP) equations \cite{ferreira_ultra-light_2021}
\begin{align}
    i\frac{\partial}{\partial t}\psi &= -\frac{1}{2m_\varphi}\nabla^2\psi+m_\varphi\Phi\psi,\label{eq:sp_1}\\
    \nabla^2\Phi &= 4\pi \rho,\label{eq:sp_2}\,,
\end{align}
where the function $\Phi = \Phi(\vec{r},t)$ represents the gravitational potential generated by the density field associated to the wavefunction, $\rho = m_\varphi |\psi|^2$. Thus, the SP equations determine the dynamics of a FDM halo. The latter is sustained by an additional internal quantum pressure arising from Heisenberg's uncertainty principle on the de Broglie wavelength scale ($\lambda_\textrm{deB}$), thus precluding system formation below $\lambda_\textrm{deB}$. N-body numerical simulations of FDM halos based on Eqs. \eqref{eq:sp_1}-\eqref{eq:sp_2} have been used to derive the physical properties of stable configurations \cite{schive_cosmic_2014, Schive2014b}. Such studies predict the generation of coherent standing waves of DM in the centers of gravitationally bound systems. These distinctive DM solitonic cores are characterized by surrounding wave interference patterns, with their azimuthal average following a NFW-like  mass density profile. The radial profile of the solitonic core is well approximated by \cite{schive_cosmic_2014, Mocz2017}
\begin{equation}
    \rho_s(r) = \rho_0\left[1+c_1\left(\frac{r}{r_c}\right)^2\right]^{-8}\,,
    \label{eq:soliton}
\end{equation}
where $c_1 = \sqrt[8]{2}-1\approx 0.091$, and $\rho_0$ corresponds to the central density of the halo
\begin{equation}\label{eq:scaling}
    \rho_0 = 0.019\left(\frac{m_\varphi}{10^{-22}\;\textrm{eV}}\right)^{-2}\left(\frac{r_c}{\textrm{kpc}}\right)^{-4} \frac{M_\odot}{\textrm{pc}^3}\,.
\end{equation}
Here, $r_c$ is a core radius parameter, and corresponds to the radial coordinate at which the density has dropped to $\sim 50\%$ of the central value $\rho_0$. Although the solitonic profile in equation \eqref{eq:soliton} seems to depend on the two parameters $m_\varphi$ and $r_c$, cosmological simulations were used to point out the following scaling relation \cite{Schive2014b}
\begin{equation}
    r_c = 1.6\left(\frac{m_\varphi}{10^{-22}\textrm{ eV}}\right)^{-1}\left(\frac{M_\textrm{halo}}{10^9\,M_\odot}\right)^{-1/3}\textrm{ kpc}.
    \label{eq:scaling_relation}
\end{equation}

The scaling relation can also be obtained from thermodynamical arguments \cite{Davies2020}, and its validity has been thoroughly investigated in \cite{Chan2022}. More specifically, a sizeable dispersion in the core halo mass relation is found for higher halo masses, pointing towards heavier solitons, especially in non-cosmological, fully virialized settings \cite{Mocz2017}.

\section{Methodology and mock data}\label{sect:methodology}

Having briefly introduced the theoretical framework of FDM in Section \ref{sect:fdm}, we can now focus on generating a dataset of star's positions and velocities mimicking a future astrometric {\em Theia}-like mission \cite{Theia2017,Malbet2019,Malbet2021}. {\em Theia}-like mission, whose main objective is the study of the dark matter properties and distribution through high accuracy measurement of the proper motions and three-dimensional positions, is designed to achieve a 100 e-fold improvements on the observational  uncertainties on proper motions with respect to {\em Gaia} satellite. Finally, a {\em Theia}-like mission is for example expected to solve the cuspy-core problem through an accurate determination of the dark matter density profile, and thus provide new insights on the intrinsic nature of dark matter.

To create a mock catalog of a target dSph galaxy, we assume spherical symmetry, which gives us information on how the positions and velocities are distributed. Then, we assume that the stars are spatially distributed in the galaxy following a Plummer profile \cite{Plummer} which is entirely characterized by its scale parameter $a$ and total stellar mass $M_*$. Finally, we assume that the velocity anisotropy parameter $\beta$ appearing in the Jeans equation is a constant, and we draw the velocity field by computing its second moments through the Jeans analysis, while the first moments are fixed by the assumption of spherical symmetry.

\subsection{Stellar Distribution Function}

The Plummer model has been used several times in literature to describe the stellar distribution in dSph galaxies (see for example \cite{walker_universal_2010,Chen2017,Massari2020}). Recently, in \cite{Massari2020}, the authors adopt a Plummer profile with best-fit parameters given in \cite{walker_universal_2010} to fit  the velocity dispersion based on the proper motions of 149 stars measured by {\em Gaia}. Additionally, in \cite{Vitral2024}, the Plummer profile is also adopted to analyse precise proper motions for hundreds of stars coming from Hubble Space Telescope observations for the Draco dSph galaxy spanning 18 years.  
Therefore, following those recent work, we choose to retain the same approach regarding the stellar distribution and, we model a dSph galaxy whose stellar distribution function follows a Plummer profile \cite{Plummer}.

We essentially consider a stellar mass distribution that generates a softened Keplerian potential. If one considers a gravitational potential of the form:
\begin{equation}
    \Phi(r) = -\frac{G M_*}{\sqrt{r^2+a^2}},
\end{equation}
$M_*$ is the total stellar mass and $a$ is the scale length, then through the Poisson equation one gets the stellar mass density profile:
\begin{equation}
    \rho(r) = \frac{3 M_*}{4\pi a^3}\left(1+\frac{r^2}{a^2} \right)^{-\frac{5}{2}}.
\end{equation}

According to the previous equation, the stellar number density is \cite{Plummer,walker_universal_2010}
\begin{equation}\label{eq:numberdensity}
    n(r)dr \sim \frac{r^2}{\left(1+\frac{r^2}{a^2} \right)^{\frac{5}{2}}} dr\,,
\end{equation}
that is the distribution of the number of particles between radii $r$ and $r+dr$, which is determined once the scale length $a$, and the total number of stars, $N_{*}$ are given.

\subsection{The mass density distribution of the dark matter halo}

The total mass distribution of a galaxy is given by both the baryonic and the DM content. However, for the case of dSph galaxies, the total mass distribution is dominated by the DM content.

As discussed in Section \ref{sect:fdm}, numerical simulations of N-body FDM predict that DM halos have a solitonic inner core surrounded by a mass density profile similar to NFW in the outer region with a transition scale on the order of 2.5 times the radius of the core \cite{schive_cosmic_2014}. Therefore, one can understand the total mass density distribution as the solitonic core, whose fitting formula is given by Equation \eqref{eq:soliton}, plus the outer NFW-like mass density distribution. This, however, adds an extra parameter that characterizes the radius at which transition from one distribution to the other happens. We define it to be proportional to the solitonic core radius: $r_t=\alpha r_c$. Therefore, the total mass density distribution can be written as
    \begin{equation}\label{eq:massdensity}
        \rho_{DM}(r)=\begin{cases}
            \rho_s(r) & \text{for } r< r_t \\[0.2cm]
             \dfrac{\rho_0}{\dfrac{r}{r_s}\left(1+\dfrac{r}{r_s}\right)^2} & \text{for } r\geq r_t, \\
        \end{cases}
    \end{equation}
where $\rho_0$ is automatically set by forcing $\rho(r)$ to be continuous at $r_t$ leaving  as free parameters that entirely characterize the mass distribution of the galaxy, $m_\varphi$, $r_c$, $\alpha$ and $r_s$.

\subsection{The velocity distribution and the Jeans analysis}

Starting from the collisionless Boltzmann equation:
\begin{equation}
    \label{Boltzmann}
    \partial_t f(\textbf{x},\textbf{v},t) + \dot{\textbf{r}} \frac{\partial f(\textbf{x},\textbf{v},t)}{\partial \textbf{r}} + \dot{\textbf{v}}\frac{\partial f(\textbf{x},\textbf{v},t)}{\partial \textbf{v}} = 0,
\end{equation}
where $f(\textbf{x},\textbf{v},t)$ is the distribution function, one can write that $\dot v =\frac{\partial \Phi}{\partial \textbf x}$ because particles interact through the gravitational force. Marginalizing $f(\textbf{x},\textbf{v},t)$ over $\textbf{v}$, one obtains a continuity equation. Instead, if one multiplies the first equation by $v_j$, then one obtains the Jeans equation
\begin{equation}
        \partial_t (n \langle v_i \rangle) + \sum_j \frac{\partial}{\partial x_j} (n\langle v_i v_j \rangle) = -n\frac{\partial\Phi(\textbf{x})}{\partial x_i},\label{eq:Jeans3D}
\end{equation}
where $\Phi(\textbf{x})$ is the gravitational potential due to the DM mass distribution. Here, $\langle...\rangle$ corresponds to the average of the quantity in parentheses. Because of the spherical symmetry the following relations hold (for more details see appendix A in \cite{deMartino2022}):
\begin{align*}
   & \langle v_r \rangle = \langle v_\theta \rangle = \langle v_\phi \rangle = 0\,\\
   & \langle v_r v_\theta \rangle=\langle v_r v_\phi \rangle=\langle v_\phi v_\theta \rangle=0\,\\
   & \langle v^2_\phi\rangle =\langle v^2_\theta\rangle.
\end{align*} 
Hence, the Equation \eqref{eq:Jeans3D} leads to the radial component of the Jeans equation:
\begin{equation}
    \frac{\partial}{\partial r} (n \langle v_r^2 \rangle) + \frac{n}{r} \beta = - n \frac{d\Phi}{dr},
\end{equation}
where we have defined
\begin{equation}
    \label{BetaDef}
\beta\equiv \beta(r) = 1 - \frac{\langle{{{v}}_{\theta}^2}\rangle(r) + \langle{{{v}}_{\phi}^2}\rangle(r)}{2\langle{{{v}}_{r}^2}\rangle(r)} =  1 - \frac{\langle{{v}}_{t}^2\rangle(r)}{\langle{{v}}_{r}^2\rangle(r)}.
\end{equation}
The anisotropy parameter will be considered, hereby, a constant parameter to infer, {\em i.e.} independent of the position. Finally, for a spherical system,  the gravitational potential $\Phi$ depends on the total enclosed mass, therefore, the solution to the radial component of the Jeans equation is given by
\begin{equation}
    \label{Jeans}
    \langle v_r^2 \rangle(r) = \frac{G r^{-2\beta}}{n(r)} \int_r^\infty dr' r'^{2\beta-2} n(r') M(r'),
\end{equation}
where $M(r')$ is the mass enclosed within the radius $r'$ and determined by the mass density profile in Equation \eqref{eq:massdensity}.

Following \cite{deMartino2022}, we will sample the velocity field from a three-dimensional multivariate normal distribution which we express in Cartesian coordinates:
\begin{equation} \label{MVNormal}
    p({\bf v}| {\boldsymbol x}) = \frac{\exp \left\{ -\frac{1}{2}\left[ {\bf v} - {\boldsymbol \langle {\bf v} \rangle({\boldsymbol x})} \right]^{T} {\bf \Sigma^{-1}({\boldsymbol x})}\left[ {\bf v} - {\boldsymbol \langle {\bf v} \rangle({\boldsymbol x})} \right]\right\} }{\sqrt{\left( 2\pi \right)^3\left| {\bf \Sigma({\bf x})}\right|}}\,,
\end{equation}
where the correlation matrix $\Sigma$ is  
\begin{align}\label{eq:cov}
&{\Sigma}({\boldsymbol x}) = \nonumber\\
&\begin{pmatrix}
\langle{{v}}_{x}^{2}\rangle - \langle{{v}}_{x}\rangle^{2} & \langle v_{xy}^{2}\rangle - \langle{{v}}_{x}\rangle\langle{{v}}_{y}\rangle & \langle v_{xz}^{2}\rangle - \langle{{v}}_{x}\rangle\langle{v}_{z}\rangle\\
 \langle{v_{xy}^{2}}\rangle - \langle{{{v}}_{x}}\rangle \langle{{{v}}_{y}}\rangle & \langle{{{v}}_{y}^{2}}\rangle - \langle{{{v}}_{y}}\rangle^{2} & \langle{v_{yz}^{2}}\rangle - \langle{{{v}}_{y}}\rangle\langle{{{v}}_{z}}\rangle \\
 \langle{v_{xz}^{2}}\rangle - \langle{{{v}}_{x}}\rangle ~\langle{{{v}}_{z}}\rangle & \langle{v_{yz}^{2}}\rangle - \langle{{{v}}_{y}}\rangle\langle{{{v}}_{z}}\rangle & \langle{{{v}}_{z}^{2}}\rangle - \langle{{{v}}_{z}}\rangle^{2}
\end{pmatrix}.
\end{align}
Therefore, adopting the number density distribution in Equation \eqref{eq:numberdensity}, and the mass density distribution in Equation \eqref{eq:massdensity}, which we need to compute the enclosed mass that generates the gravitation potential, once the anisotropy parameter is given, the covariance matrix is also determined (after converting the solution of the Jeans equation from the spherical to the Cartesian coordinates) and the velocity field can be sampled from the probability distribution in Equation \eqref{MVNormal}.  

\begin{table*}[!ht]
\setlength{\tabcolsep}{10pt}
\renewcommand{\arraystretch}{1.5}
\centering
\begin{tabular}{llccr}
\hline 
\textbf{Parameter} & \textbf{Description} &  \textbf{Input Value}  &  \textbf{Unit}            &  \\
\hline
$N_*$ & Total number of stars sampled & [100, 1000, 2000, 4000, 6000] & - & Known\\
$a$       & Scale Factor for the Plummer Distribution & 0.296 & kpc   & Known              \\
$m_\varphi$  & Mass of FDM particle                      & $10^{-22}$ & eV  & Inferred           \\
$r_c$     & Scale parameter of the Solitonic profile  & $ 0.98$ &kpc  & Inferred           \\
$\beta$   & Anisotropy parameter                      & [0.25, 0.0, -0.25]   &   -    &  Inferred           \\
$\alpha$  & Transition parameter between the profiles & 2.1       & -  & Inferred           \\
$r_s$     & Scale parameter for the NFW profile       & 3   & kpc     & Inferred  \\
$\sigma_v$ &  Instrumental error on the velocity components       & [0.0, 1.0, 3.0]   & km/s     & Known \\
\hline
\end{tabular}
\caption{Summary of relevant parameters in the model.}\label{tab:input_params}
\end{table*}

\subsection{Sampling the mock catalogs}

To create a mock catalog based on the theoretical model introduced in the previous sections, we need to fix a set of parameters, some of which will be inferred later to estimate the precision achievable in the framework of future {\it Theia}-like astrometric missions. As target galaxy, we have chosen the dSph galaxy Draco because it is supposed to be highly dominated by DM showing a mass-to-light ratio $M/L_V= 300 M_\odot/L_\odot$ \cite{Lokas2005}  and, therefore, a very good laboratory to test DM models. The total number of stars $N_*$ that are supposed to be observed will be set to different values ranging from 100 to 6000 stars. The Plummer scale parameter is set to $0.296$ kpc as reported in \cite{walker_universal_2010}. The boson mass of the FDM particle will be set to $m_\varphi = 10^{-22}$ eV, with a core radius of $0.98$ kpc. Based on the scaling relation in equation \eqref{eq:scaling_relation} this will return a total mass of $4.3 \times 10^9 M_\odot$, which is comparable to the total mass estimated through observations for our target galaxy \cite{walker_universal_2010}. The anisotropy parameter $\beta$ will be set to three different values to evaluate whether having a radial, isotropic, or tangential velocity distribution would affect the capability to infer the right model parameters. And, finally, the transition radius between the inner and outer mass density distribution of the DM matter halo will be set to {a reference value of} $2.1r_c$, and the scale radius of the outer DM mass density distribution will be set to $3$ kpc. All these parameters are summarized in Table \ref{tab:input_params} together with a description and stating whether they will be inferred or not.

First, we have to sample the radial position of $N_*$ stars from the probability distribution given in Equation \eqref{eq:numberdensity}. Then, we will assign the angular coordinates $(\theta,\phi)$ by uniformly sampling the following distributions: $\cos \theta \in (-1,1)$ and $\phi \in (0,2\pi)$. By construction, the system of reference is centered on the center of mass of the target galaxy. Once the positions of the stars are given, we can solve the Jeans equation \eqref{Jeans} to estimate the three-dimensional velocity dispersion of the sample. Therefore, the covariance matrix in Equation \eqref{eq:cov} is determined and we can sample the velocities from the multivariate distribution in Equation \eqref{MVNormal}, after the required change of coordinates from spherical to Cartesian.
\begin{figure}[!ht]
    \centering
    \includegraphics[width=\columnwidth]{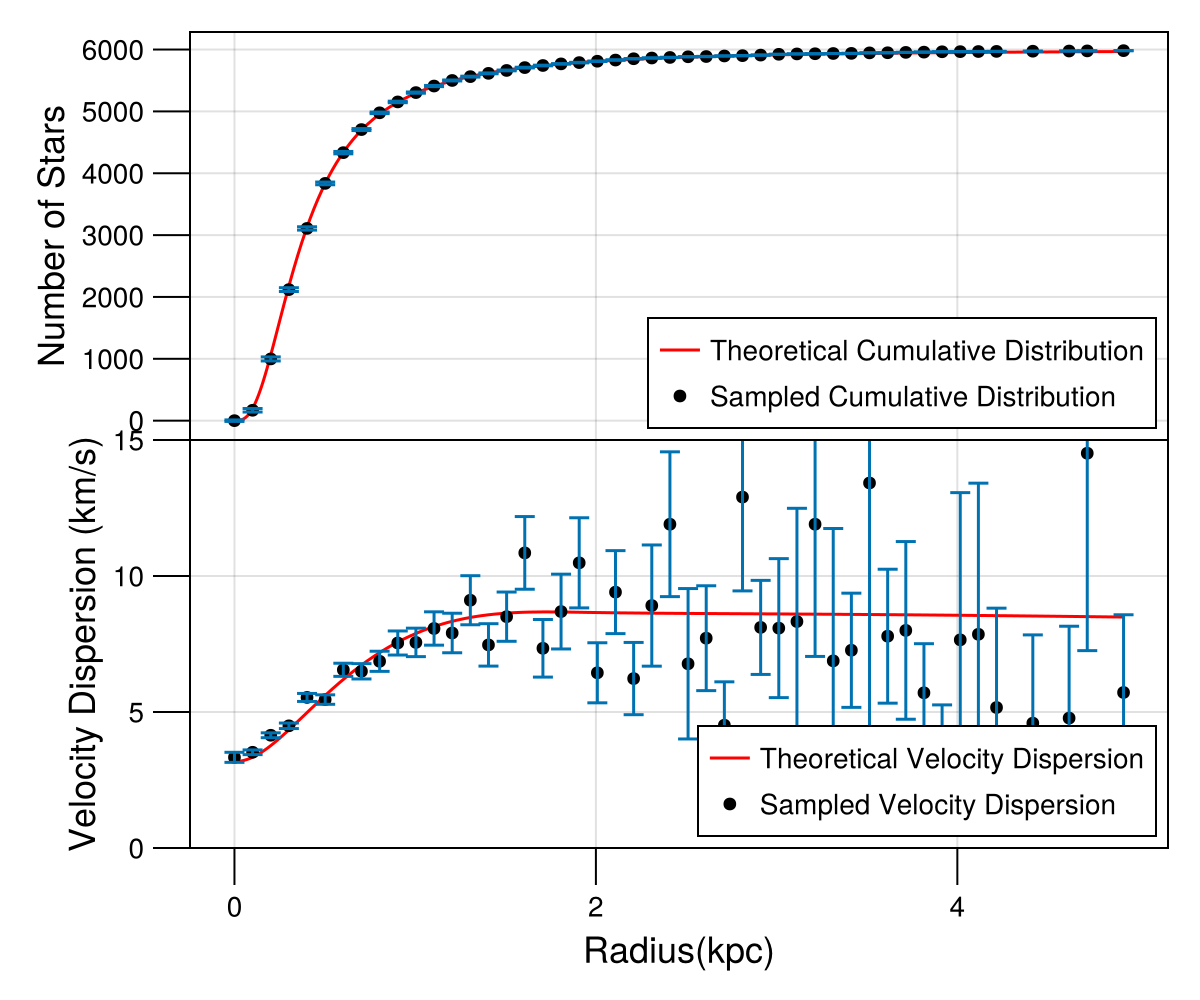}
    \caption{Comparison between sampled and theoretical values for the cumulative number distribution and velocity dispersions. The Figure shows the results for the mock catalog containing $N_*=6000$ stars and obtained by setting $\beta=0$. {\it Upper panel}: The 
cumulative number of stars expected from a Plummer distribution (red solid line) is compared with the cumulative number of stars computed according to equation~\eqref{Number}.  {\it { Lower} panel}: The radial component of the velocity dispersion profile obtained by solving the Jeans equation~\eqref{Jeans} (red solid line) is compared to the one obtained by the mock data according to the equation~\eqref{JeansDiscrete} (black points).   }
    \label{fig:NcumDisp}
\end{figure}

To illustrate the goodness of the sampling procedure, we show in the upper panel of Figure \ref{fig:NcumDisp} the cumulative number of stars expected from a Plummer distribution (solid red line), and in the lower panel of Figure~\ref{fig:NcumDisp}, we show the expected radial component of the velocity dispersion, which is obtained by solving equation~\eqref{Jeans}, as the solid red line, and compare it with the velocity dispersion of the sample. Following \cite{deMartino2022}, the cumulative number of stars {theoretically} expected from a Plummer distribution (solid red line in the upper panel) is given by
\begin{equation}\label{Number}
N_{\rm th}(<r) = \frac{N_* r^3}{(r^2 + a^2)^{3/2}}\,.
\end{equation} 
The discrete cumulative number of stars (black dots in the upper panel) is obtained by computing the number of stars in each equally spaced {radial} bin ($N_b$). To each point is associated an error bar, which, however, is smaller than the point size. Finally, the discrete velocity dispersion (black dots in the lower panel) of the sample which is computed as follows
\begin{equation}\label{JeansDiscrete}
\langle{{{v}}_{r,k}^2(r)}\rangle^{1/2} = \left[\sum_i^{N_b} (v_r)_i^2/N_b\right]^{1/2}\, .
\end{equation}
where $(v_r)_i$ is the radial velocity of the $i$-th star in the $k$-th radial bin, and the error bars in each bin are compute as follows
\begin{equation}
        \sigma_{\langle{{{v}}_{r,k}^2}\rangle^{1/2}}= \left[\frac{\langle{{{v}}_{r,k}^2}\rangle}{2(N_{b}-1)}\right]^{1/2}\,.
\end{equation}
For both the cumulative number of stars and the velocity dispersion, the results obtained from our sampling procedure and the theoretically expected ones are within shot-noise fluctuations. Finally, we checked that we obtain self-consistent results for each mock catalogue that we have generated. We also checked that changing halo parameters does not alter the generation of the mock catalogues.

\subsection{Inferring the model parameters}

The mock catalogs list the positions and velocities of $N_*$ stars. The parameters we are going to infer are $\mathbf \Theta = (\log_{10}(m_\varphi/eV), r_c, \beta, \alpha, r_s)$ as reported in Table \ref{tab:input_params}. Thus, the likelihood is
\begin{equation*}
    p(d |\Theta) = \prod_{i=1}^N p(d_i|\Theta),
\end{equation*}
where we have denote the data of the $i$-th star as $\mathbf d_i = (\bf r_i, \bf v_i)$. Therefore, motivated by Equation \eqref{MVNormal}, the $i$-th likelihood is given by
\begin{equation}
    p(r_i, {\bf v_i}|\Theta) =  \frac{\exp \left\{ -\frac{1}{2}\left[ {\bf v} - { \langle {\bf v} \rangle({ r_i})} \right]^{T} { C^{-1}({r_i})}\left[ {\bf v} - { \langle {\bf v} \rangle({r_i})} \right]\right\} }{\sqrt{\left( 2\pi \right)^3\left| { C({r_i})}\right|}}\,,
\end{equation}
where the covariance matrix is $C (r) = \Sigma(r) $ if instrumental errors are not taken into account, and $C(r) = \Sigma(r) + S(r)  $ on the contrary. Assuming that the measurements are not correlated in the three directions, the covariance matrix $S(r)$ is diagonal and its non-zero elements are the squared errors of the error measurements. We assume no error in the positions. Our choice of priors is specified in Table \ref{Priors}, while the instrumental error in the measurements is set to $\sigma_{{\rm{v}}_{i}}= [1,3,5]$ km/s. 

To explore the parameter space, we employ a variant of Hamiltonian Monte Carlo (HMC) algorithm, called No U-Turn Sampling (NUTS) \cite{NUTS}.  We ran eight chains in parallel, each sampling 2000 points with 250 of them being used for burn-in. To ensure convergence, we checked that the Monte Carlo Standard Error (MCSE) of each parameter is less than 5\% of the marginal posterior standard deviation, that the Effective Sample Size (ESS) of both bulk and tail is greater than 100 times the number of chains, and that the Gelman-Rubin statistic ($\hat{R}$) is less than 1.1. All of these diagnostic methods are described in \cite{ConvergenceOfMCMC}.

\begin{table}
\setlength{\tabcolsep}{15pt}
\renewcommand{\arraystretch}{1.5}
\centering
\begin{tabular}{cc}
\hline
\textbf{Parameters}      &          \textbf{Priors}         \\
\hline
$\log_{10}(m_\varphi/eV)$    & U(-24,-17) \\
$r_c$         & U(0.4,4)   \\
$\beta$                & U(-5,0.99) \\
$\alpha$                & U(2,10)    \\
$r_s$           & U(2,10)   \\
\hline
\end{tabular}
\caption{Uniform prior distribution on the model parameters.}
\label{Priors}
\end{table}

\section{Results and discussions}\label{sect:results}

\begin{table*}[!ht]
\resizebox{\textwidth}{!}{\begin{tblr}{
  cells = {c},
  cell{2}{1} = {r=4}{},
  cell{2}{2} = {r=4}{},
  cell{6}{1} = {r=4}{},
  cell{6}{2} = {r=4}{},
  cell{10}{1} = {r=4}{},
  cell{10}{2} = {r=4}{},
  cell{14}{1} = {r=4}{},
  cell{14}{2} = {r=4}{},
  cell{18}{1} = {r=4}{},
  cell{18}{2} = {r=4}{},
  vline{4} = {-}{},
  hline{1-2,6,10,14,18,22} = {-}{},
}
\textbf{Parameter}            & \textbf{Input Value} & ${\bm \sigma_v(km/s)}$ & ${\bm N_*=100}$                  & ${\bm N_*=1000}$                 & ${\bm N_*=2000}$                 & ${\bm N_*=4000}$                   & ${\bm N_*=6000}$                    \\

$\log_{10}(m_\varphi/\textrm{eV})$ & $-22.0$     & 0                & $-21.95^{+0.16}_{-0.14}$ & $-22.02^{+0.05}_{-0.05}$ & $-21.99^{+0.03}_{-0.03}$ & $-22.00^{+0.02}_{-0.02}$   & $-22.01^{+0.02}_{-0.02}$    \\

                     &             & 1                & $-21.96^{+0.15}_{-0.14}$ & $-22.04^{+0.05}_{-0.05}$ & $-22.01^{+0.03}_{-0.03}$  & $-22.02^{+0.02}_{-0.02}$   & $-22.03^{+0.02}_{-0.02}$    \\
                     
                     &             & 3          & $-21.99^{+0.15}_{-0.15}$ &$-22.07^{+0.05}_{-0.05}$ & $-22.03^{+0.03}_{-0.04}$ & $-22.07^{+0.03}_{-0.03}$&$-22.06^{+0.02}_{-0.02}$\\
                     
                     &             & 5                & $-22.00^{+0.15}_{-0.15}$ & $-22.10^{+0.05}_{-0.05}$ & $-22.06^{+0.04}_{-0.04}$ & $-22.07^{+0.03}_{-0.03}$   & $-22.09^{+0.02}_{-0.02}$    \\

$r_c$ (kpc)          & 0.98        & 0                & $0.91^{+0.19}_{-0.17}$   & $1.00^{+0.07}_{-0.04}$   & $0.97^{+0.04}_{-0.04}$   & $1.00^{+0.03}_{-0.03}$     & $1.00^{+0.03}_{-0.03}$      \\

                     &             & 1                & $0.93^{+0.19}_{-0.18}$   & $1.04^{+0.07}_{-0.07}$   & $1.00^{+0.04}_{-0.04}$   & $1.03^{+0.03}_{-0.03}$     & $1.03^{+0.03}_{-0.03}$   \\
                     & & 3 & $0.98^{+0.22}_{-0.18}$ & $1.11^{+0.07}_{-0.07}$ & $1.07^{+0.05}_{-0.05}$ &$1.16^{+0.04}_{-0.04}$ & $1.11^{+0.03}_{-0.03}$ \\
                     
                     &             & 5                & $1.04^{+0.24}_{-0.20}$   & $1.17^{+0.05}_{-0.05}$   & $1.14^{+0.06}_{-0.05}$   & $1.16^{+0.04}_{-0.04}$     & $1.19^{+0.04}_{-0.04}$      \\

$\beta$              & 0.0         & 0                & $-0.07^{+0.15}_{-0.16}$  & $-0.06^{+0.04}_{-0.04}$  & $-0.01^{+0.03}_{-0.03}$  & $0.01^{+0.02}_{-0.02}$     & $0.00^{+0.02}_{-0.02}$   \\

                     &             & 1                & $-0.09^{+0.16}_{-0.19}$  & $-0.08^{+0.05}_{-0.05}$  & $-0.03^{+0.03}_{-0.03}$  & $0.00^{+0.02}_{-0.02}$ & $-0.00^{+0.02}_{-0.02}$ \\
                     
                     & & 3 & $-0.14^{+0.18}_{-0.21}$&$-0.14^{+0.06}_{-0.06}$& $-0.07^{+0.04}_{-0.04}$ & $-0.08^{+0.03}_{-0.03}$ &$-0.05^{+0.02}_{-0.02}$\\
                     
                     &             & 5                & $-0.21^{+0.21}_{-0.26}$  & $-0.21^{+0.07}_{-0.07}$  & $-0.11^{+0.05}_{-0.05}$  & $-0.08^{+0.03}_{-0.03}$    & $-0.10^{+0.03}_{-0.03}$     \\

$\alpha$             & 2.1         & 0                & $4^{+4.0}_{-2.0}$            & $2.50^{+1.9}_{-0.3}$     & $2.45^{+0.20}_{-0.19}$    & $2.33^{+0.15}_{-0.15}$     & $2.15^{+0.14}_{-0.10}$      \\

                     &             & 1                & $4^{+4.0}_{-2.0}$            & $3.0^{+4}_{-0.1}$        & $2.47^{+0.19}_{-0.20}$     & $2.35^{+0.15}_{-0.15}$     & $2.20^{+0.15}_{-0.12}$      \\
                     & & 3 & $5^{+3.0}_{-3.0}$ & $3.39^{+4.40}_{-1.02}$ & $2.51^{+0.27}_{-0.21}$ & $2.45^{+0.19}_{-0.18}$ & $2.31^{0.19}_{0.17}$\\
                     &             & 5                & $5.4^{+3.1}_{-2.9}$      & $5.0^{+3.0}_{-2.1}$      & $2.6^{+0.4}_{-0.3}$      & $2.45^{+0.19}_{-0.18}$     & $2.44^{+0.25}_{-0.21}$      \\

$r_s$ (kpc)           & 3.0         & 0                & $6.1^{+2.7}_{-2.7}$      & $6.8^{+2.6}_{-2.8}$      & $7.0^{+2.1}_{-2.8}$      & $6.3^{+2.4}_{-2.6}$        & $5.5^{+2.7}_{-2.2}$         \\

                     &             & 1                & $6.1^{+2.7}_{-2.7}$      & $6.2^{+2.6}_{-2.8}$      & $6.9^{+2.1}_{-2.8}$      & $6.4^{+2.1}_{-2.8}$        & $5.7^{+2.7}_{-2.2}$         \\
                     & & 3 & $6.0^{+2.7}_{-2.6}$ &$6.1^{+2.6}_{-2.8}$ & $6.7^{+2.3}_{-2.7}$ & $6.3^{+2.3}_{-2.7}$& $5.8^{+2.8}_{-2.5}$\\
                     
                     &             & 5                & $6.0^{+2.7}_{-2.7}$      &$6.0^{+2.7}_{-2.7}$      & $6.6^{+2.4}_{-2.9}$      & $6.3^{+2.6}_{-2.7}$ & $6.1^{+2.6}_{-2.7}$         
\end{tblr}}
\caption{The first column reports the parameters of the model and the second column lists the fiducial value of those parameters. The third column shows the uncertainty on the measure of each velocity component of the stars in the mock catalogs assumed in the statistical analysis. The other columns report the median and 68\% confidence level of the parameters of the model estimated through the MCMC analysis, and $N_*$ represent the number of stars in the mock catalogue. The results shown in the Table illustrate the case of an isotropic velocity field, $\beta=0$.}\label{tab:results}
\end{table*}

\begin{figure*}[!ht]
    \centering
    \includegraphics[width=2\columnwidth]{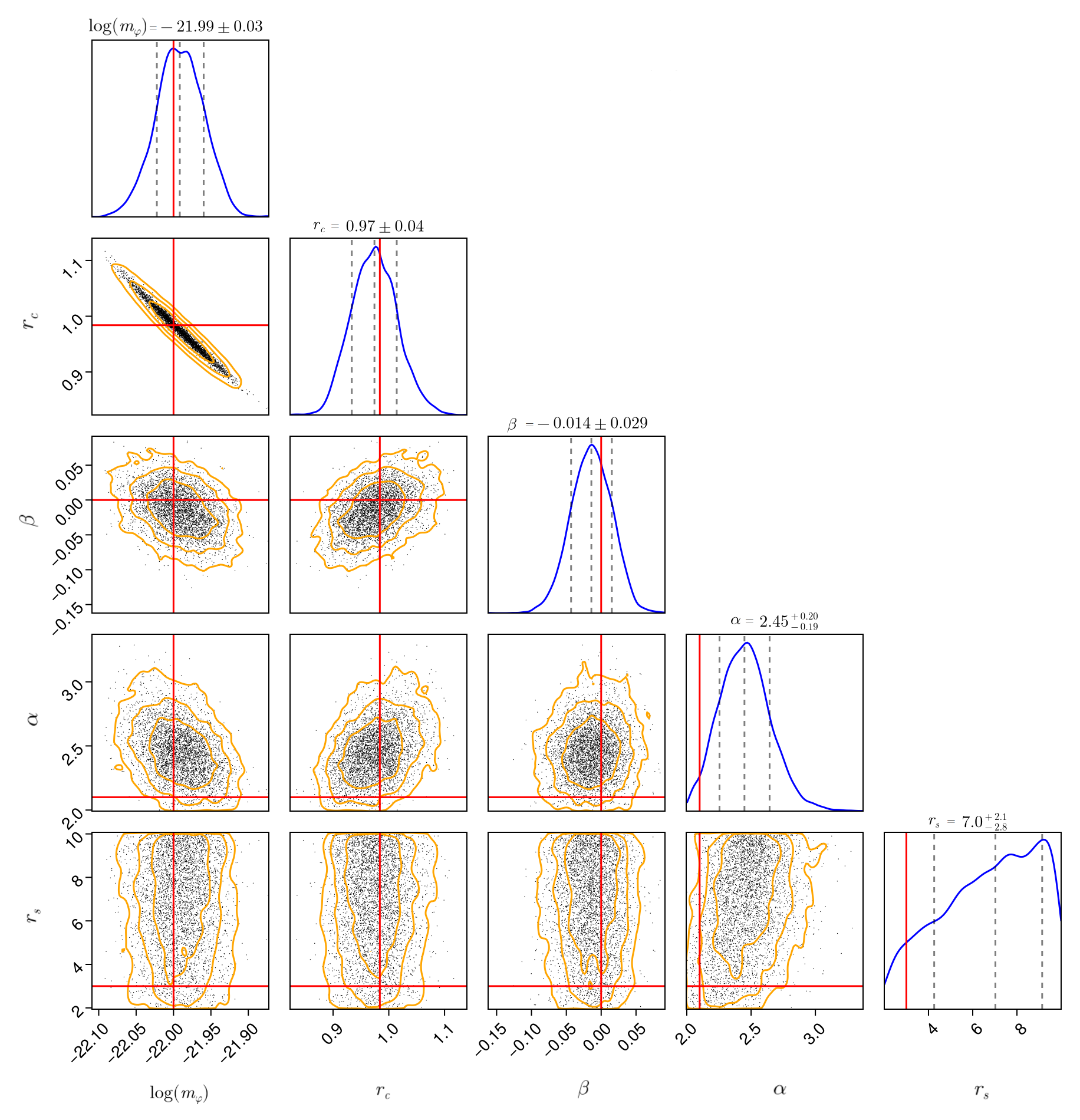}
    \caption{The corner plot shows the posterior distributions of the model parameters. On top of each column, the 1D posterior distribution is shown (blue line) and the median and statistical uncertainty at the 68\% confidence level are reported  for the corresponding parameter, which can be compared with the input value represented by a vertical red line. {In the off-diagonal plots,} the 2D density distribution of the MCMC chains is shown for each couple of parameters. In these panels, the input values (red lines) and the confidence levels at 68\%, 95\%, and 99\% are also depicted (orange contours).}
\label{fig:corner}.
\end{figure*}

In this section, we show and discuss the results of the MCMC analysis carried out using our mock catalogs. We will use as a representative case the results obtained for the target galaxy assuming an isotropic velocity field, $\beta=0$, and using the mock catalog containing $N_*=2000$ stars. Then, we will discuss the constraining power shown by the proper motion on the boson mass, and the core and transition radii, as well as we will establish the impact of the instrumental uncertainty on the measure of each velocity component of the stars on the final constraints. All results are reported in Figure \ref{fig:corner} and Table \ref{tab:results}. 

\begin{figure*}[!ht]
    \centering
    \includegraphics[width=0.85\columnwidth]{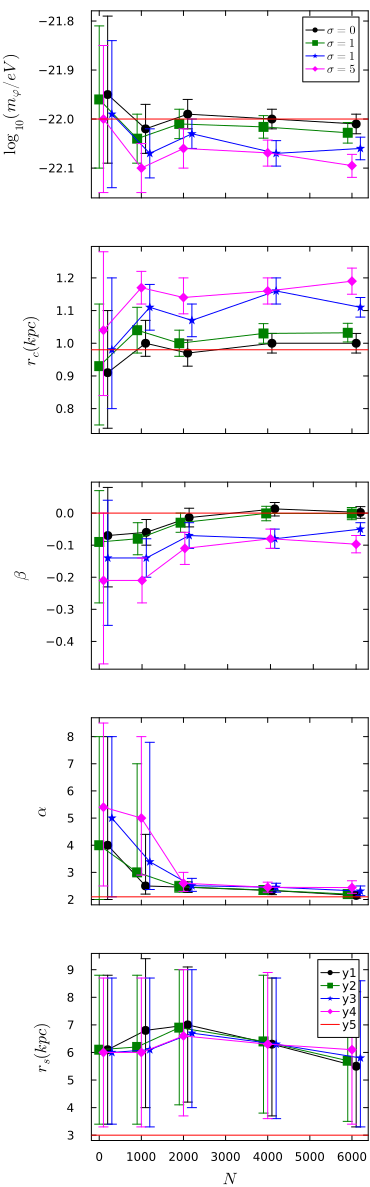}
    \caption{The Figure depicts how the median and 68\% uncertainty of the model parameters change with the number of stars. In the first panel, there are depicted the results obtained from the MCMC for the boson mass $\log_{10}(m_\varphi/eV)$, in the second panel is shown the core radius $r_c$, in the third panel is depicted the anisotropy parameter $\beta$, in the forth panel is shown the transition parameter $\alpha$ and, finally, in the last panel is reported the scale radius $r_s$. The red lines represent the input values, while the black circles, the green squares, the blue stars and the magenta diamonds represent the median of the parameters with the corresponding 68\% uncertainty obtained when a measurement error of 0, 1, 3, and 5 km/s is assumed, respectively. }
    \label{fig:errors}
\end{figure*}

Figure \ref{fig:corner} reports the posterior distributions of the model parameters $\mathbf \Theta = (\log_{10}(m_\varphi/\textrm{eV}), r_c,\beta, \alpha, r_s)$, and the 2D density plots of each couple of parameters  with confidence contours (orange lines) at 68\%, 95\%, and 99\% of the confidence level. The red lines represent the input parameters (which are listed in the second column of Table \ref{tab:results}). On top of each column, we report the median value and the statistical uncertainty at 68\% confidence level for each parameter. The input values of the boson mass $\log_{10}(m_\varphi/eV)$, the core radius $r_c$, and the anisotropy parameter $\beta$ are recovered within the 68\% confidence interval, while the input value of the transition parameter $\alpha$ is recovered within the 95\% confidence interval. Our methodology is therefore unbiased on those parameters if the proper motion of at least 2000 stars is measured. However, we have to remark that the scale radius $r_s$ remains unconstrained, independently of the number of stars contained in the mock catalog (not even in the best-case scenario with 6000 stars that we have simulated). This can be easily understood noting that the number of stars, and hence of the proper motion measurements, decrease with the radius leading to a less precise measurement of the velocity dispersion (as shown in the lower panel of Figure \ref{fig:NcumDisp}. Those results do not change when we introduce an uncertainty $\sigma_v$ in the measure of each velocity component of the stars. As shown in Table \ref{tab:results}, increasing the uncertainty to $\sigma_v=1$ km/s does not change anything, while setting $\sigma_v=3$ km/s slightly degrades the final results. Indeed, in this case, the anisotropy parameter is also recovered within the 95\% confidence interval, and it is still unbiased anyway. Moreover, increasing the number of stars does not improve the result. Finally, Table \ref{tab:results} also shows that having $\sigma_v>3$ km/s introduces a bias. In fact, setting $\sigma_v=5$ km/s will affect our ability to detect the solitonic structure of the DM halo. Those results are also visually shown in Figure \ref{fig:errors}. Starting from the top panel, we show how the median and 68\% uncertainty of the model parameters $\mathbf \theta = (\log_{10}(m_\varphi/eV), r_c,\beta, \alpha, r_s)$ change with the number of stars. The red lines represent the input values, while the black circles, the green squares, the blue stars, and the magenta diamonds represent the median of the parameters with the corresponding 68\% uncertainty obtained when a measurement error of 0, 1, 3, and 5 km/s is assumed, respectively. Finally, we checked that setting a radially or tangentially biased velocity field would not affect the final results, as shown in Table \ref{tab:results2}.

\begin{table}[!ht]
\begin{tblr}{
  cells = {c},
  cell{2}{1} = {r=3}{},
  cell{2}{2} = {r=3}{},
  cell{5}{1} = {r=3}{},
  cell{5}{2} = {r=3}{},
  cell{8}{1} = {r=3}{},
  cell{8}{2} = {r=3}{},
  cell{11}{1} = {r=3}{},
  cell{11}{2} = {r=3}{},
  cell{14}{1} = {r=3}{},
  cell{14}{2} = {r=3}{},
  vline{4} = {-}{},
  hline{1-2,5,8,11,14,17} = {-}{},
}
\textbf{Parameter}   & \textbf{Input Value} & \textbf{Input for~$\beta$} & $\bm N=2000$             \\
$\log_{10}(m_\varphi/eV)$ & $-22.0$              & -0.25                      & $-21.99^{+0.03}_{-0.03}$ \\
                     &                      & 0.0                       & $-21.99^{+0.03}_{-0.03}$ \\
                     &                      & 0.25                       & $-22.01^{+0.04}_{-0.04}$ \\
$r_c (kpc)$          & 0.98                 & -0.25                      & $0.97^{+0.04}_{-0.04}$   \\
                     &                      & 0.0                       & $0.97^{+0.04}_{-0.04}$   \\
                     &                      & 0.25                       & $1.00^{+0.05}_{-0.05}$   \\
$\beta$              & -                    & -0.25                      & $-0.22^{+0.04}_{-0.04}$  \\
                     &                      & 0.0                          & $-0.01^{+0.03}_{-0.03}$  \\
                     &                      & 0.25                       & $0.26^{+0.02}_{-0.02}$   \\
$\alpha$             & 2.1                  & -0.25                      & $2.14^{+0.14}_{-0.09}$   \\
                     &                      & 0.0                          & $2.45^{+0.20}_{-0.19}$    \\
                     &                      & 0.25                       & $2.21^{+0.24}_{-0.15}$   \\
$r_s(kpc)$           & 3.0                  & -0.25                      & $6.8^{+2.1}_{-2.5}$      \\
                     &                      & 0.0                         & $7.0^{+2.1}_{-2.8}$      \\
                     &                      & 0.25                       & $6.3^{+2.5}_{-2.7}$      
\end{tblr}
\caption{The same of Table \ref{tab:results} with different values of the anisotropy parameters. We have considered an input model with the measurement error set to $\sigma_v = 0$ km/s, and a mock catalogue containing 2000 stars.}\label{tab:results2}
\end{table}

In the case of the isotropic velocity field,  using mock observations of 2000 stars and assuming an uncertainty on the measurements of the velocity components below 1 km/s, we will be able to constrain the mass of the boson and the core radius with an accuracy of roughly 3\%, while the transition parameter with an accuracy of $\sim 8\%$, as reported in the Table  \ref{tab:results}, hence, detecting the whole solitonic structure of the DM halo. On the other hand, as mentioned above, we cannot constrain the outermost structure of the halo parametrized by the scale radius due to the low number of observed stars expected in those regions. Increasing the number of observed velocities to 6000 will improve the accuracy on $m_\varphi$ and $r_c$ to $\sim 2\%$ and the accuracy on the transition radius to the $\sim 7\%$. On the other hand, if the error in the measured velocity components is greater than $1$ km / s, the precision on $m_\varphi$, $r_c$, and $\alpha$ is reduced to 4\%, 5\% and 13\%, respectively, for $\sigma_v =5$ km / s. Finally, Table \ref{tab:results2} shows that similar accuracies can be achieved also in the cases of radially and tangentially biased velocity field.

\section{Conclusions}\label{sect:conclusions}

Despite the enormous amount of evidence on the gravitational effects that DM produces, its fundamental nature is still unknown \cite{DeMartino2020,Salucci2021} and is known to lie in the physics beyond the Standard Model of particles. The most promising candidate has been for a long while the Weak Interacting Massive Particles (or WIMPs) whose existence is still not confirmed by the experimental detection. Due to the lack of experimental evidence, many other candidates have been proposed and, among them, one of the most promising is represented by FDM. These are ultralight bosonic particles whose cosmological evolution is governed by the SP equations \eqref{eq:sp_1} and \eqref{eq:sp_2}. Cosmological simulations have shown that these particles form a DM core in the center of each virialized halo, and that such a core is surrounded by quantum interference patterns on the de
Broglie scale, whose azimuthal average follows an NFW-like mass density profile \cite{schive_cosmic_2014}. This DM candidate is capable of addressing several small-scale challenges faced by the CDM model \cite{DeMartino2020}, but it also suffers from some inconsistencies that need to be clarified. Indeed, while galactic tests based on the kinematics of the stars within the bulge of the Milky Way, and of the dSph and ultra-diffuse galaxies
point to a boson with a mass of the order of $\sim10^{-22}$ eV \cite{Chen2017,Broadhurst2020,DeMartino2020PDU,Pozo2020,Pozo2021,Pozo2023}, cosmological analysis based on the Lyman-$\alpha$ forest set lower bounds on the boson mass $\gtrsim 2\times10^{-20}$ eV \cite{Armengaud2017,Rogers2021} at the 95\% confidence level. Therefore, there is an important need to design new tests to constrain the mass of the boson.
One possibility is represented by the timing of pulsars near the Galactic center \cite{DeMartino2017}, which should become available with the forthcoming radio interferometer SKA. Another possibility that we have explored here is to use the proper motion of stars in galaxies instead of projected dispersion velocity or circular velocity to distinguish between a boson of mass of the order of $\sim 10^{-22}$ eV and of the order of $\sim 10^{-20}$ eV.

As a first step, we created mock catalogs simulating the measurement of the proper motions of $N_*$ stars in a Draco-like dwarf galaxy. As mentioned in Section \ref{sect:methodology}, the galaxy's DM halo is constructed with an ultralight boson of mass $10^{-22}$ eV, and a core radius of 0.98 kpc. Finally, stars are distributed following a Plummer profile, and the velocity components of each star are sampled from the multivariate Gaussian distribution given in the equation. \eqref{MVNormal}. Then, we have used an MCMC algorithm to determine the accuracy to which next generation {\it Theia}-like astrometric satellites will be capable of constraining the model parameters. All results are summarized in Tables \ref{tab:results} and \ref{tab:results2}. We have found that we will need to measure the proper motions of at least 2000 stars with an instrumental error $\sigma_v \leq 3$ km/s on the velocity components to constrain the solitonic structure with 3\% accuracy. Furthermore, the transition between the soliton and the NFW-like density profile would be detected with a relative uncertainty of 7\%. This would undoubtedly help to achieve two objectives: first, to determine the boson mass on galactic scales, probing the whole velocity field and constraining the boson mass with unprecedented accuracy; and, second, to shed light on the need of having a multiple axion-like particles with a broad mass range and naturally arising in the String Axiverse to explain the different values of the boson mass on galactic and cosmological scales. For comparison, the Jeans analysis of the current velocity dispersion data in the Draco galaxies presented in \cite{Chen2017} constrained the boson mass with an uncertainty of $\Delta \log_{10}(m_\varphi/eV) \approx 0.46$, while we demonstrate that with proper motion of at least 2000 stars  we might improve more than one order of magnitude the precision on the boson mass reaching $\Delta \log_{10}(m_\varphi/eV) \approx 0.03$  and fitting also the transition radius as the same time. Other analyses based on the velocity dispersion along the line of sight have been shown in \cite{Morales2017} leading to results similar to \cite{Chen2017}. Conversely, using a sample of 18  ultra-faint
dwarf galaxies, the boson mass has been constrained with a much larger uncertainty of \cite{Hayashi2021} due to the introduction of a radial dependent anisotropy parameter that strongly degraded the constraining power of the data by adding additional free parameter.

Finally, we would like to discuss several important assumptions behind those results. The first is the assumption of spherical symmetry for our synthetic dwarf galaxy, although it is well known that real dwarfs appear elliptical on the sky \cite{Irwin1995,Salomon2015}. This assumption is rather simplistic and extending to axisymmetric system adopting the axisymmetric Jeans equations would add more degrees of freedom to the problem, actually degrading the final results. The second one is the assumption $\beta=\textrm{const.}$ rather than a radially-dependent anisotropy parameter \cite{Okoli2016}. However, a radially dependent $\beta$ would add at least one more parameter to constrain, leading to degraded results. Additionally, in \cite{deMartino2023}, it was shown that the tension on the boson mass between cosmological and galactic probes cannot be alleviated introducing a variable anisotropy parameter. Therefore, since our aim is to prove the capability of future proper motion measurements to recover the structure of a FDM halo, we decided to rely on the simpler assumption, leaving the issue of introducing a radially dependent $\beta$ to a future work. Finally, the last assumption is the use of the Plummer profile. Here we are not interested in recovering also the stellar distribution. However, including the stellar distribution in the statistical algorithm to constrain its parameters would clearly add degeneracies and degrade the final results. A better strategy would be to fit the stellar distribution from the luminosity profile and, with those results, to fit the DM distribution from the velocity field. Finally, future research avenues should also focus on incorporating more realistic observational uncertainties, exploring non-spherical models, or considering a radially-dependent anisotropy parameter, to estimate the impact of the different assumptions on the final results.

\section*{Acknowledgments}
IDM and RDM acknowledge financial support from the grant PID2021-122938NB-I00 funded by MCIN/AEI/10.13039/501100011033. RDM also acknowledges support from Consejeria de Educación de la Junta de Castilla y León and the European Social Fund.  IDM also acknowledges support from the grant SA097P24 funded by Junta de Castilla y León and by “ERDF A way of making Europe”.
\clearpage

\bibliographystyle{apsrev4-2}
\bibliography{Biblio.bib}

\end{document}